\documentstyle[preprint,aps,epsfig]{revtex}
\begin{document} 
  
\normalsize 

\title{Chemical Equilibration in Pb+Pb collisions at the SPS}

\author{P. Braun-Munzinger$^1$, I. Heppe$^2$, J. Stachel$^2$}
\address{
$^1$ Gesellschaft f\"ur Schwerionenforschung, D 64291 Darmstadt,
Germany\\
$^2$ Physikalisches Institut der Universit\"at Heidelberg, D 69120
Heidelberg, Germany }
 
\date{\today} 

\maketitle

\begin{abstract}

An improved statistical model with excluded volume corrections and
resonance decays is introduced and applied to the 
complete presently available set of particle ratios as measured by the
various experiments at the SPS in Pb+Pb collisions. The results imply
that a high degree of hadrochemical equilibration is reached at chemical
freeze-out  in Pb+Pb collisions.

\end{abstract} 
\narrowtext

Heavy ion collisions at ultra-relativistic energies are studied to look
for signs of the production of a
quark-gluon plasma phase which subsequently hadronizes.
In this context one of the crucial questions is whether thermal
and chemical equilibrium is achieved at some stage of the
collision. Applying a statistical model which assumes equilibrium, and
testing experimental data against model predictions is one way of
testing reality against a thermally and chemically equilibrated fireball
at the point of hadro-chemical freeze-out.

The present statistical model -- like its predecessor which was presented
in \cite{AGS,SPS} -- is based on the use of a grand canonical ensemble
to describe the partition function and hence the density of the
particles of  species \(i\) in an equilibrated fireball:

\begin{equation}
n_i= \frac{g_i}{2 \pi^2} \int_0^\infty \frac{p^2 \, {\rm
d}p}{e^{(E_i(p)-\mu_i)/T} \pm 1}
\label{grundgl}
\end{equation}

\noindent with particle density \(n_i\), spin degeneracy \(g_i\),
\(\hbar\) = c = 1, momentum \(p\), total energy \(E\)  and chemical
potential \(\mu_i = \mu_B B_i-\mu_S S_i-\mu_{I_3} I^3_i\). The
quantities \( B_i\), \( S_i\) and \( I^3_i\) are the baryon,
strangeness and three-component of the isospin quantum numbers of the
particle of species \(i\). The temperature T and the baryochemical
potential \(\mu_B\) are the two independent parameters of the model,
while the volume of the fireball V, the strangeness chemical potential
\(\mu_S\), and the isospin chemical potential \(\mu_{I_3}\) are fixed
by the three conservation laws \footnote{These conservation laws apply
strictly only for quantities which are evaluated over the complete
phase space.} for

\begin{eqnarray}
\mbox{baryon number:} &\quad& V \sum_i n_i B_i = Z+N, \\
\mbox{strangeness:} &\quad& V \sum_i n_i S_i = 0,  \\
\mbox{and charge:} &\quad& V \sum_i n_i I^3_i = \frac{Z-N}{2}.
\end{eqnarray}
Here, Z and N are the proton and neutron numbers of the colliding nuclei.
The hadronic mass spectrum used in the calculations extends over all
mesons with masses below 1.5 GeV and baryons with masses below 2
GeV. This limits the temperature up to which thermal model calculations
are trustworthy to T$_{max} < 185$ MeV. We note, however, that
calculations with higher temperatures should anyway be considered with
caution as the mass spectrum for heavier hadrons is not sufficiently
well known. Maybe somewhat unexpectedly  the
neutron excess in Pb plays only a minor role in the determination of
particle ratios.

To take into account a more realistic equation of state we incorporate the
repulsive interaction at short distances between hadrons by means of an
excluded volume correction.  A number of different corrections have been
discussed in the literature. Here we choose that proposed in
\cite{exclv1,exclv2}:

\begin{equation}
 p^{excl.} (T,\mu)= p^{id.gas}(T,\hat{\mu}); \qquad \mbox{with }
\hat{\mu} = \mu - v_{eigen} \; p^{excl.}(T,\mu). 
\label{druck}
\end{equation}

This thermodynamically consistent approach to simulate interactions
between particles by assigning an eigenvolume \(v_{eigen}\) to all
particles, modifies the pressure \(p\) within the
fireball. Equation \ref{druck} is recursive, as it uses the modified chemical
potential \(\hat{\mu}\) to calculate the pressure, while this pressure
is also used in the modified chemical potential, and the final value is found by iteration. Particle densities are
calculated by substituting \(\mu\) in eq. \ref{grundgl} by the
modified chemical potential \(\hat{\mu}\).

The eigenvolume has to be chosen appropriately to simulate the repulsive
interactions between hadrons, and we have investigated the consequences
for a wide range of parameters for this eigenvolume in
\cite{diplomarbeit}. Note that the eigenvolume is $v_{eigen} = 4
\frac{4}{3} \pi R^3$ for a hadron with radius R. Assigning the same
eigenvolume to all particles can reduce particle densities drastically
but hardly influences particle ratios.  Ratios may differ strongly,
however, if different values for the eigenvolume are used for different
particle species.

\begin{figure}[h]
\centerline{\epsfig{figure=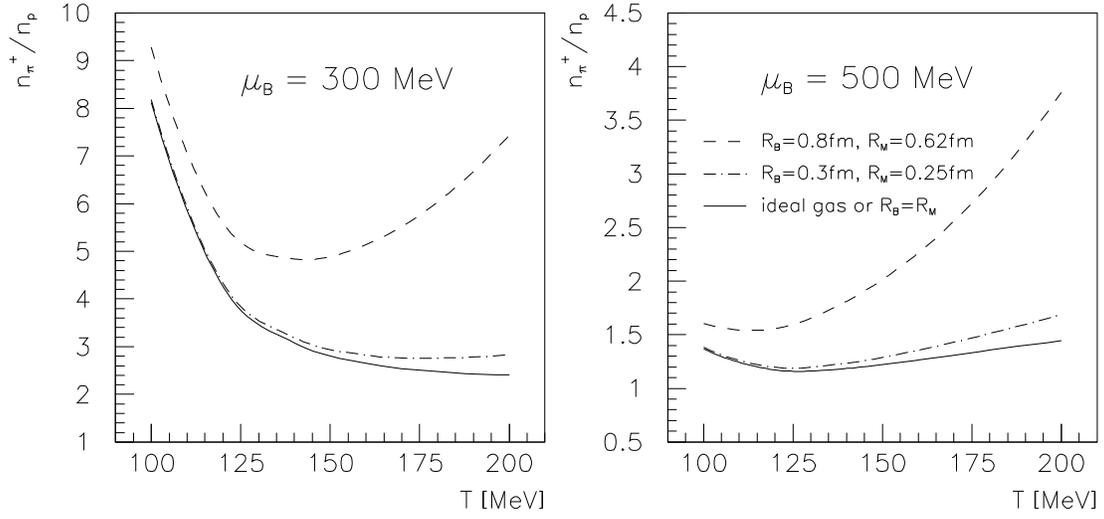,width=14.5cm}}

\vspace{.5cm}

\caption{The influence of different excluded volume corrections for baryons and
mesons on the \(\rm{\pi^+/p}\) ratio.} 
\label{exclratios}
\end{figure}

Two different scenarios were explored: First, we chose the radius of all
baryons according to the charge radius of the proton, which lies at 0.8
fm, and assigned a smaller radius of 0.62 fm to all mesons, as suggested
in \cite{exclv2}. This drastic correction reduces the thermally produced
particle density in Pb+Pb collisions by a factor of seven as compared to
the ideal gas case. Furthermore, because baryons take up more space than
mesons, their creation is suppressed in favor of meson production. Hence
the meson to baryon ratio increases strongly. This is illustrated in
Figure \ref{exclratios} where we plot, for different chemical
potentials, the temperature dependence of the pion/nucleon ratio and
compare it to predictions from the ideal gas scenario.

However, the nucleon-nucleon or pion-pion interaction is not repulsive
at such large distances. A more physical approach is to determine, for
nucleons, the eigenvolume according to the hard-core volume known from
nucleon-nucleon scattering \cite{bohrmott}.  Consequently, we assigned
0.3 fm as radius for all baryons. For mesons we expect the eigenvolume
not to exceed that of baryons.  Therefore, to illustrate the effect, we
kept the ratio of meson to baryon radii as above, implying a meson 
radius of 0.25 fm. As can be seen in Figure \ref{exclratios}, the
resulting particle ratios are much closer to predictions using an ideal
gas scenario;  absolute yields are reduced by about 30 \%. In the
absence of detailed information about the meson-meson interaction at
short distances we assumed for the following calculations that \(\rm
R_{baryon}=R_{meson}\)=0.3 fm.

After thermal ``production'', resonances and heavier particles are
allowed to decay, therefore contributing to the final particle yield of
lighter mesons and baryons. Decay cascades, where particles decay in
several steps, are also included. A systematic parameter regulates the
amount of decay products resulting from weak decays. This allows to
simulate the different reconstruction efficiencies for particles from
weak decays in different experiments.
 
This model is now applied to Pb+Pb collisions at maximum SPS energy.
We have used all data currently available \footnote{The
$\overline{\Lambda}/\Lambda$ ratio from the NA49 collaboration will be
revised (P. Seyboth, NA49 collaboration, private communication) and is
therefore not included in Table \ref{table}. For the same reason we
have replaced the $\Xi^-/\Lambda$ ratio from NA49 by the ratio $(\Xi^+
+ \Xi^-)/(\overline{\Lambda} + \Lambda)$.}. We adjust the free parameters
T and \(\mu_B\) such that they reproduce best all particle ratios
available at the moment. We did not include in the fit the \(
2\phi/(\pi^+ + \pi^-)\) ratio, because, for this ratio, the currently
available two experimental values exhibit a rather large discrepancy.
The results for the best $\chi^2$ (see below) are shown in Table
\ref{table}. For technical reasons only one of the
\(\rm{\overline{p}/p}\) ratios (the NA49 value) was included in the
fit.  All experimental particle ratios are taken from data integrated
over transverse momentum and integrated over rapidity y to the extent
data are available as shown in Table \ref{table}.  Using this
procedure strongly reduces the possible influence on particle ratios
of dynamical effects such as hydrodynamic flow or particle production
from a superposition of fireballs \cite{cley_jaipur}.

The criterium for the best fit was either a minimum in
\begin{equation} 
\chi^2=\sum_{i} \frac{( {\cal R}_i^{\rm exp.}-{\cal R}_i^{\rm
model})^2}{\sigma_i^{\rm 2}}, 
\end{equation}
or a minimum in the quadratic deviation
\begin{equation} 
q^2=\sum_{i} \frac{( {\cal R}_i^{\rm exp.}-{\cal R}_i^{\rm
model})^{\rm 2}}{ ({\cal R}_i^{\rm model})^{\rm 2}}.
\end{equation}

In the above equations \( {\cal R}_i^{\rm model}\) and \({\cal
  R}_i^{\rm exp.}\) are the \(i\)th particle ratio as calculated from
our model or measured in the experiment, and \(\sigma_i\) represent
the errors in the experimental data points as quoted in the
experimental publications. We have used both the $\chi^2$ and the
quadratic deviation measure to estimate the influence of possible
systematic errors which are generally not included in the data. The
deviation between these two analyses gives an indication of the
accuracy of the parameters extracted from this data set.

\begin{figure}[h]
\centerline{\epsfig{figure=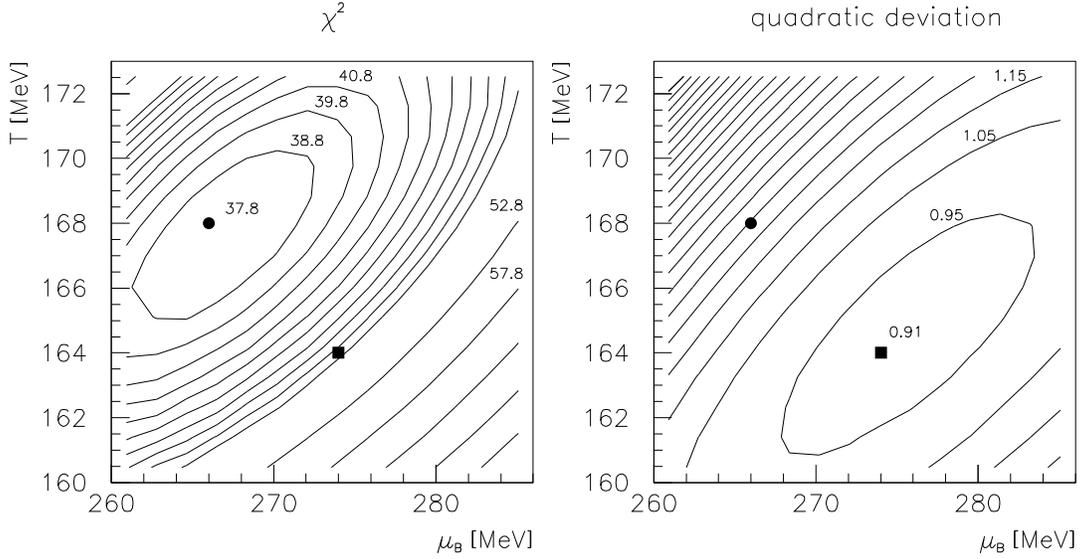,width=14.5cm}}

\vspace{.5cm}

\caption{\(\chi^2\) and quadratic deviation for the comparison between
model particle ratios and data explored over a wide range of
parameters. The dot represents the parameter set with minimum
\(\chi^2\), the square the set with minimum quadratic deviation.}
\label{contours}
\end{figure}

As one can see from Figure \ref{contours}, the best fit in terms of
$\chi^2$ was achieved at T = 168 \(\pm\) 2.4 MeV, \(\mu_B\) = 266
\(\pm\) 5 MeV, with \(\mu_S=71.1\) MeV and \(\mu_{I_3}=-5.0\) MeV.
The minimal quadratic deviation is found at T = 164 MeV, \(\mu_B\) =
274 MeV. These small differences give an indication of the systematic
uncertainties of the procedure.

\begin{figure}[h]
\centerline{\epsfig{figure=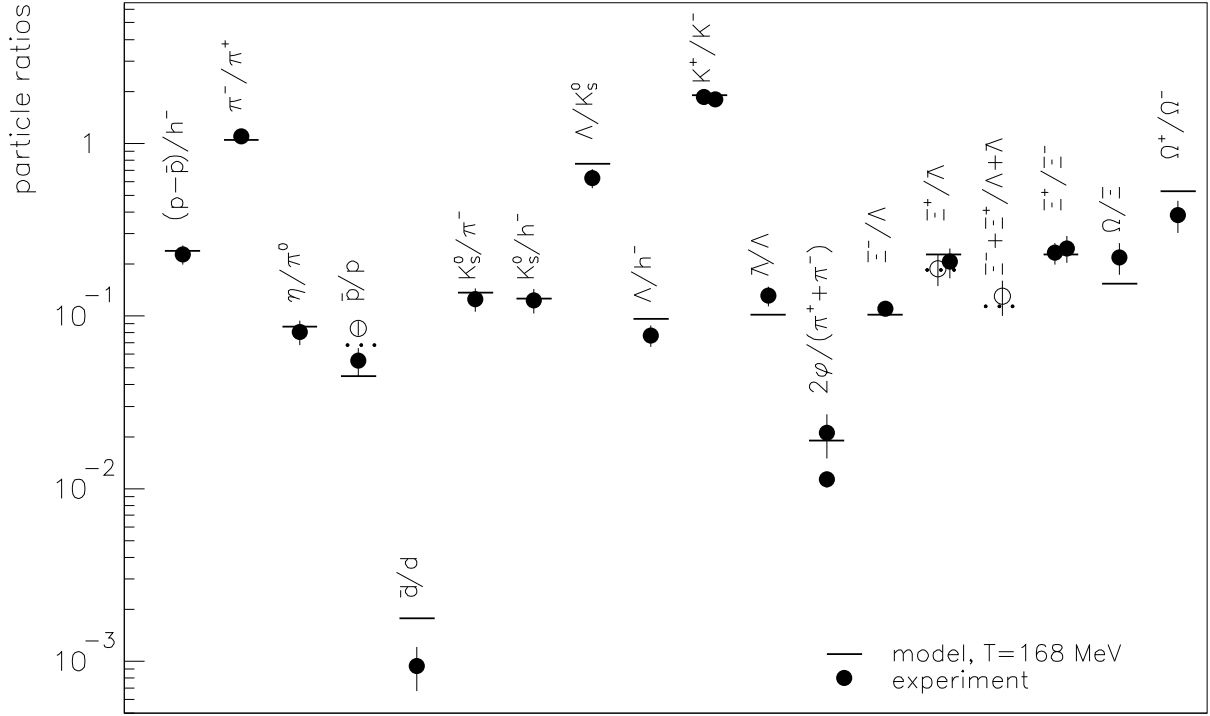,width=18.0cm}}
\caption{Comparison between model and experimental particle
ratios. For experimental data, errors and information about feeding
corrections see Table 1 and references there.}
\label{fool}
\end{figure}

The overall agreement between model and data is quite good, as can be
seen in Figure \ref{fool} and Table \ref{table}. Choosing a slightly
different excluded volume correction with \(\rm R_{baryon}\)=0.3 fm
and \(\rm R_{meson}\)=0.25 fm yields very similar results.  Using
significantly larger eigenvolumes leads to much poorer
agreement. Comparison between data and the model, e.g., with \(\rm
R_{baryon}\)=0.8 fm and \(\rm R_{meson}\)=0.62 fm yields
\(\chi^2_{min}\)=180, which is roughly 5 times as large as the
value shown in Figure \ref{contours}. In any case, we have excluded such
large radii for independent physics reasons as discussed above.

Furthermore, using such large radii leads to an eigenvolume of all
particles which would occupy 55\% of the total volume and could
therefore not be considered a ``correction''. The total fireball volume
would increase to roughly 20000 fm\(^3\),  exceeding even the
fireball volume estimated using pion interferometry. As discussed in
\cite {harry,stachel}, the total fireball volume in central Pb+Pb
collisions at thermal freeze-out should be about 13500 fm\(^3\).

In the small excluded volume scenario with R\(\rm _{baryon}\)=R\(\rm
_{meson}\)=0.3 fm the fireball volume of 2800 fm\(^3\) is considerably
smaller. The corresponding pion density is then 0.60 pions/fm\(^3\),
significantly exceeding the measured pion density of roughly 0.12
pions/fm\(^3\) \cite{stachel}. This is not surprising, however, as the
calculated pion density of 0.6/fm$^3$ is determined at chemical freeze-out
corresponding to T=168 MeV. If one lets this fireball expand
isentropically to 125 MeV, the temperature roughly corresponding to
thermal freeze-out as indicated by particle spectra \cite{pbmst}
and two-pion correlations \cite{harry} the corresponding pion density is
0.084/fm$^3$, close to the experimental value. Note that the calculated
pion density would increase further by about 30 \% 
if one were to reduce to zero the meson radius in the excluded volume
correction.

The present results, in particular those involving multi-strange
baryons, imply that no separate strangeness suppression factor is needed
to describe the available Pb+Pb data at SPS energy. In fact, the mean
value of the experimental to calculated yields ratios involving
$\Delta$S = 1, i.e. those which are sensitive to a possible  overall
strangeness suppression, is 0.96 $\pm$ 0.05, consistent with unity. This
conclusion 
differs from that reached in a recent investigation \cite{gaz} where,
however, only a very resticted set of ratios was used for comparison
with thermal model predictions. An interesting anomaly would arise if
the $\phi$-meson yield converges to the low value reported by the
NA49 collaboration (see Table \ref{table}), since this meson carries two
units of hidden strangeness. To reconcile this with the results by the
WA97 collaboration on cascade or omega-baryon production would be a challenge.

We further note that the improved model discussed here was also applied
to the AGS data collected in \cite{AGS}.  The best fit, obtained for
R\(\rm _{baryon}\)=R\(\rm_{meson}\)=0.3 fm, yields T = 125 (+3 - 6) MeV
and $\mu_B$ = 540 $\pm 7$ MeV, well in line with the calculations
reported in \cite{AGS}. Here, the corresponding $\pi^+$ and proton
densities of 0.051/fm$^3$ and 0.053/fm$^3$ agree well with those
estimated from particle interferometry \cite{agsplb,nu} (0.058/fm$^3$
and 0.063/fm$^3$, respectively) implying that, at AGS energy, thermal
and chemical freeze-out take place at nearly identical temperatures.

The good agreement between the predictions of the thermal model and the
measured particle ratios implies that thermal and chemical equilibrium
is established (or at least closely approached) in the fireball at
hadrochemical freeze-out. Furthermore, it is interesting to note that
the resulting temperature and chemical potential values are very close
to where we believe is the  phase boundary  between hadronic matter and the
quark-gluon plasma  \cite{qgp}. It is, therefore,  quite probable that
the system crosses this phase boundary shortly before it freezes out
hadrochemically.

\begin{table}[H]

\caption{Experimental particle ratios compared to model predictions with
${\rm R_{baryon}=R_{meson}=0.3}$ fm, T=168 MeV, $\mu_B$=266 MeV,
$\mu_S$=71.1 MeV, $\mu_{I_3}$=-5.0 MeV. \\
$^{(a)}$ : feeding from weak decays excluded,\\
$^{(b)}$ : feeding from weak decays included,\\
$^{(c)}$ : cuts exclude feeding of $\Lambda$ from $\Sigma^\pm$ and $\Xi$ . \\ 
In all three cases feeding in the model was tuned accordingly. In all
other cases feeding from weak decays is assumed to be 50\%.}

\label{table}

\newpage

\begin{tabular}{|l|l|l|lll|}
\hline &&&&&\\ & model & exp. data & exp. & y-range & ref. \\ &&&&&\\
\hline \hline 
\(\rm (p-\overline{p})/h^-\) & 0.238 & 0.228(29)
$^{(a)}$ & NA49 & 0.2-5.8 & \cite{NA49,jones}\\ 
\(\rm \overline{p}/p\)
& 0.045 & 0.055(10) $^{(a)}$ & NA44 & 2.3-2.9 & \cite{NA44}\\ 
\(\rm \overline{p}/p\) & 0.060 & 0.085(8) $^{(b)}$ & NA49 & 2.5-3.3 &
\cite{jones}\\ 
\(\rm \overline{d}/d\) & 1.78 \(10^{-3}\) & 0.94(27)
\(10^{-3}\) & NA44 & midrapidity & \cite{bearden} \\ 
\(\rm \pi^-/\pi^+ \) & 1.05 & 1.1(1) & NA49 & all & \cite{joerg}\\ 
\(\rm \eta/\pi^0\) &
0.087 & 0.081(13) & WA98 & 2.3-2.9 & \cite{peitz}\\ 
\(\rm K^0_s/\pi^-\) & 0.137 & 0.125(19) & NA49 & all & \cite{hirschegg}\\
\(\rm K^0_s/h^-\) & 0.126 & 0.123(20) & WA97 & 2.4 - 3.4 &\cite{WA97}\\ 
\(\rm \Lambda/h^-\) & 0.096 & 0.077(11) & WA97 & 2.4 - 3.4 &
\cite{WA97}\\ 
\(\rm \Lambda/K^0_s\) & 0.76 & 0.63(8) & WA97 & 2.4 -
3.4 & \cite{WA97}\\ 
\(\rm K^+/K^-\) & 1.90 & 1.85(9) & NA44 & 2.4-3.5
& \cite{NA44}\\ & 1.90 & 1.8(1) & NA49 & all & \cite{NA49}\\ 
\(\rm \overline{\Lambda}/\Lambda\) & 0.102 & 0.131(17) $^{(c)}$ & WA97 & 2.4
- 3.4 & \cite{WA97}\\ 
\(\rm \Xi^-/\Lambda\) &0.102 & 0.110(10)
$^{(c)}$ & WA97 & 2.4 - 3.4 & \cite{WA97}\\ 
\(\rm \Xi^+/\overline{\Lambda}\) & 0.185 & 0.188(39) & NA49 & 3.1-4.1 &
\cite{frank}\\ & 0.228 & 0.206 (40) $^{(c)}$ & WA97 & 2.4 - 3.4 &
\cite{WA97}\\
\(\rm (\Xi^+ + \Xi^-)/(\Lambda + \overline{\Lambda})\) & 0.114 & 0.13(3) 
& NA49 & 3.1-4.1 & \cite{NA49-1}\\  
\(\rm \Xi^+/\Xi^-\) & 0.228 & 0.232(33) & NA49 &
3.1-4.1 & \cite{frank}\\ & 0.228 & 0.247(43) & WA97 & 2.4 - 3.4 &
\cite{WA97}\\ 
\(\rm \Omega^+/\Omega^-\) & 0.53 & 0.383(81) & WA97 &
2.4 - 3.4 & \cite{WA97}\\ 
\(\rm \Omega/\Xi\) & 0.154 & 0.219(45) &
WA97 & 2.4 - 3.4 & \cite{WA97}\\
\hline
\(\rm 2 \phi/(\pi^++\pi^-)\) & 19.0 \(10^{-3}\) & 21(6) \(10^{-3}\) &
NA50 & 2.9 - 3.9 & \cite{NA50}\\ 
 & 19.0 \(10^{-3}\) & 12.2(13) \(10^{-3}\) & NA49 & all & \cite{puehl}\\
\hline
\end{tabular}
\end{table}

\vspace{7.cm}

\end{document}